\documentclass[12pt]{book}

\usepackage[dvips]{graphicx,color}
\usepackage{makeidx,physics,cosmology}
\makeauthorindex
\makeindex

\BookTitle{Frontier in Astroparticle Physics and Cosmology}
\CopyRight{\copyright 2004 by Universal Academy Press, Inc.}

\begin{document}

\BookTitle{\itshape Frontier in Astroparticle Physics and Cosmology}
\CopyRight{\copyright 2004 by Universal Academy Press, Inc.}
%\tableofcontents
\pagenumbering{arabic}

\chapter{Deuterium and $ ^{7}$Li Concordance in Inhomogeneous Big Bang
         Nucleosynthesis Models}

\author{%
Juan F. LARA\\
{\it Department of Physics and Astronomy, Clemson University, 120-C Kinard
     Laboratory, Clemson, SC 29634-0978, USA}}
%
% Please note:
% One \AuthorContents{} is necessary
% for EACH CONTRIBUTION, for the contents page and
% One \AuthorIndex{} is necessary
% for EACH AUTHOR, for the index.
%
\AuthorContents{J.\ Lara} %%%%%%% <=== It is the data for CONTENTS. Please enter all author's name that should be inithialized.

\AuthorIndex{Lara}{J.} %%%%%%% <=== It is the data for AUTHOR INDEX. Please enter a author's name that should be inithialized.

\section*{Abstract}

Recent observational constraints on primodial deuterium and $ ^7$Li correspond 
to different values of the baryon-to-photon ratio when applied to the Standard 
Big Bang Nucleosynthesis ( SBBN ) model.  In this article these constraints are
applied to baryon Inhomogeneous ( IBBN ) models.  A depletion factor of 3.4 
applied to the $ ^{7}$Li constraints will bring $ ^{7}$Li, deuterium and 
$ ^{4}$He constraints in concordance for both the SBBN and IBBN models.  A 
depletion factor of 6.1 will bring concordance for the IBBN model alone.

\section{Introduction:  SBBN Models}

In Big Bang Nucleosynthesis free neutrons and protons fuse to form gradually
heavier nuclei via nuclear and weak reactions.  In the Standard BBN model the
universe is homogeneous and isotropic.  The baryon-to-photon ratio $\eta$ is
a variable in the SBBN model.

Figure (1) shows the mass fraction of $ ^{4}$He and the log of the abundance 
ratios of deuterium to hydrogen and $ ^{7}$Li to hydrogen for the SBBN case,
wherein they are functions of $\eta$.  These results can be compared with 
observations of primordial abundances.  The $ ^{4}$He mass fraction has been 
measured as low as 0.228 \cite{oss97} and as high as 0.248 \cite{it03}.  
Kirkman et al \cite{ktsol03} have measured [d/H] = 
$2.78_{-0.38}^{+0.44} \times 10^{-5}$ while Ryan et al \cite{rbofn00} have 
measured [$ ^{7}$Li/H] = $1.23_{-0.32}^{+0.68} \times 10^{-10}$.  [d/H] 
corresponds to $\eta = ( 5.6 - 6.6 ) \times 10^{-10}$ while [$ ^{7}$Li/H] to
a different range, $\eta = ( 1.6 - 4.1 ) \times 10^{-10}$.  

\begin{figure}[t]
  \begin{center}
    \includegraphics[height=22pc]{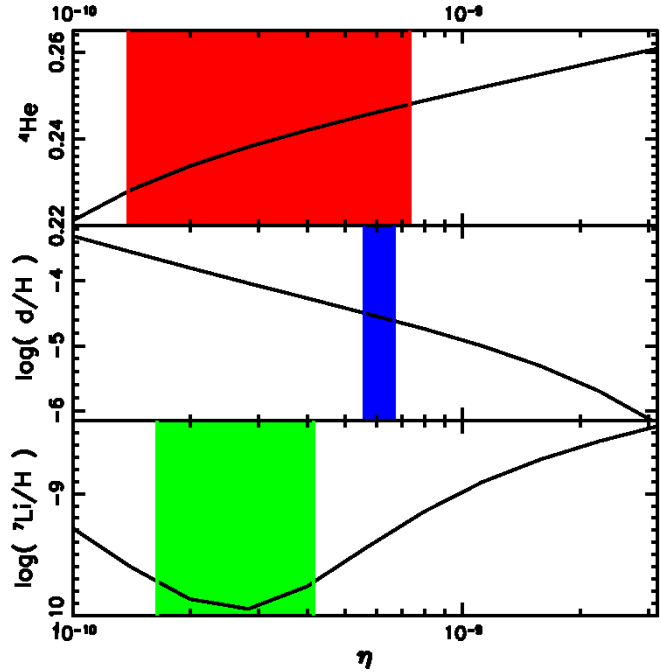}
  \end{center}
  \caption{SBBN:  $ ^{4}$He, [d/H], and [$ ^{7}$Li/H] as functions of $\eta$.
           Constraints are in color.}
\end{figure}

\section{Methods:  IBBN Models}

Certain theories of baryogenesis \cite{ks01} lead to inhomogeneous baryon 
distributions at the time of BBN.  As an example, this article uses a model 
with a low density cylindrical core and a high density cylindrical outer shell.
At temperatures $T > $ 13 GK weak reactions gradually convert neutrons to 
protons.  At some time depending on the model size neutrons diffuse into the 
low density core until they are homogeneously distributed.  The outer shell may
maintain a high proton density at $T = $ 0.9 GK, when BBN occurs.  BBN will 
occur earlier in the outer shell, depleting neutrons there.  Neutrons then back
diffuse to the outer shell and BBN occurs primarily in the high density shell.

Along with $\eta$, IBBN models will depend on $r_{i}$, the size of the model,
$R_{\rho}$, the initial ratio of high to low density, $f_{v}$, the volume 
fraction of the high density region, and the geometry of the model.  To 
simplify the model only neutrons can diffuse.

\section{Results and Discussion}

Figure (2) is an abundance contour map for the cylindrical shell model.  
$r_{i}$ is the size of the model at $T = $ 100 GK.  A region of concordance 
between [d/H] \cite{ktsol03} and the $ ^{4}$He \cite{it03} constraints is
shown in yellow.  The region marked by the [$ ^{7}$Li/H] constraints is shown
in light green.  

\begin{figure}[t]
  \begin{center}
    \includegraphics[height=25pc,angle=90]{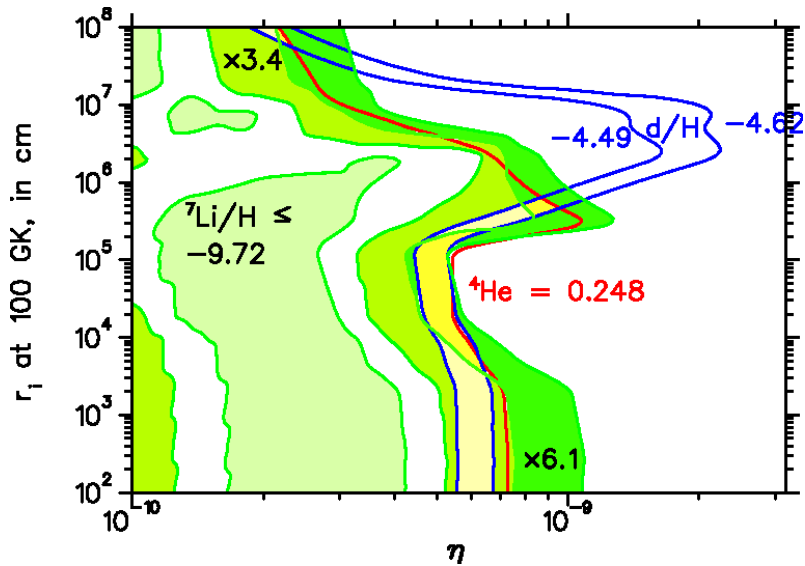}
  \end{center}
  \caption{$ ^{4}$He, [d/H], and [$ ^{7}$Li/H] constraints in the cylindrical
           shell model, as functions of $\eta$ and $r_{i}$.}
\end{figure}

The effect of inhomogeneity to the isotope abundances is comparable to a shift
in the value of $\eta$ in the standard model.  Abundance contours tend to run
parallel to another, and the region marked out by the Ryan et al \cite{rbofn00}
constraints on $ ^{7}$Li never coincides with the [d/H]-$^{4}$He region of 
concordance for any value of $r_{i}$.  The $ ^{7}$Li constraints with a 
depletion factor of 3.4 are shown in moderate green.  With this depletion 
factor the constraints agree the [d/H] and $ ^{4}$He constraints for the SBBN
case.  The constraints agree for this IBBN model as well for $r_{i}$ up to 
20000 cm, and for a region where $r_{i} = $ [ 150000, 500000 ] cm.

As $r_{i}$ increases neutron diffusion occurs at later times.  The baryon 
density of the outer shell at the time of BBN increases.  For any set value of
of $\eta$ $ ^{4}$He increases while deuterium production decreases, and the
contours of $ ^{4}$He and [d/H] shift to lower $\eta$.  Production of 
$ ^{7}$Be, which decays into $ ^{7}$Li, increases with increasing $r_{i}$ more
quickly though than the change in the other light isotopes.  For $r_{i} = $ 
[ 20000, 150000 ] cm the $ ^{7}$Li constraints need a depletion factor of 
6.1, shown in dark green, to be in concordance with the other isotope 
constraints.  

At $r_{i}$ = 600000 cm neutron diffusion starts to occur late enough to 
coincide with BBN.  Neutron diffusion to the low density core competes with 
back diffusion due to earlier BBN in the high density outer shell.  In this 
case the model produces both large amounts of $ ^{4}$He and deuterium.  The
region of concordance between $ ^{4}$He and [d/H] reaches its maximum value of
$r_{i}$ for this case.

Contour maps were made for spherical shell, condensed sphere, and condensed
cylinder geometries as well as cylindrical shell geometry.  In these maps the
shapes of the contour lines become more exaggerated if the volume fraction is
decreased.  But the contour lines maintain the same basic appearance.  In all
geometries a thin region of concordance existed between [d/H] and $ ^{4}$He
constraints but not with $ ^{7}$Li without the depletion factors mentioned 
above.

\section{Conclusions}

The existence of a region of concordance between [d/H] and $ ^{4}$He 
constraints in IBBN models is independent of the geometry of the model.  In 
IBBN models the $ ^{7}$Li constraints still need a depletion factor in order to
be in concordance with the other constraints.  A factor of 3.4 can bring all 
the constraints in concordance for both SBBN and small size IBBN models.  
Larger size IBBN models alone can satisfy the constraints with a factor of 6.1.

A future article will discuss the influence of neutron diffusion on final
abundances in much greater detail.  The cases where $ ^{7}$Be production is
greatly increased and when neutron diffusion and BBN coincide will be 
discussed as those cases determine the sizes of the concordance regions.  The
influence of other details such as proton and isotope diffusion and neutrino
degeneracy \cite{kajino02} also needs to determined.  Models for $ ^{7}$Li 
depletion will be kept in mind to compare with the depletion factors determined
in this article.

\section{List of Symbols/Nomenclature}

  $T$ = Temperature, GK \hspace{1.0in} $\eta$ = baryon to photon ratio

                   $r_{i}$ = size of the IBBN model at $T = $ 100 GK, cm

                $R_{\rho}$ = initial ratio of high to low baryon density

                   $f_{v}$ = fraction of volume that has high density.

                     [d/H] = abundance ratio of deuterium to free protons

             [$ ^{7}$Li/H] = abundance ratio of $ ^{7}$Li to free protons

%%%%%%%%%%%%%%%%%%%%%%%%%%%%%%%%%%
%% thebibliography environment %%
%%%%%%%%%%%%%%%%%%%%%%%%%%%%%%%%%

%%%%%%%%%%

\end{document}